\documentclass[10pt]{article}
\usepackage{amsmath}
\usepackage{amssymb}
\usepackage{graphicx}
\usepackage{color}
\usepackage{cite}   

\usepackage{setspace}
\makeatletter
\renewcommand{\@biblabel}[1]{\quad#1.}
\makeatother

\usepackage{bigstrut}

\newcommand{\EQ}[1]{Eq.~(\ref{eq:#1})}

\begin{document}

\begin{flushleft}
{\Large \textbf{The McDonald-Kreitman Test and its Extensions under Frequent Adaptation: Problems and Solutions}}\\

\vspace{1em}
Philipp W. Messer,
Dmitri A. Petrov\\

\vspace{1em}

Department of Biology, Stanford University, Stanford, CA 94305, USA\\

\vspace{2em}

Biological Sciences: Evolution

\vspace{2em}

Corresponding author:\\
\vspace{1em}
Philipp Messer\\
Department of Biology\\
Stanford University\\ 
371 Serra St.\\ 
Stanford, CA 94305-5020\\
Phone: 650 736 2249\\ 
messer@stanford.edu\\

\end{flushleft}

\newpage

\section*{Abstract}

Population genomic studies have shown that genetic draft and background selection can profoundly affect the genome-wide patterns of molecular variation. We performed forward simulations under realistic gene-structure and selection scenarios to investigate whether such linkage effects impinge on the ability of the McDonald-Kreitman (MK) test to infer the rate of positive selection ($\alpha$) from polymorphism and divergence data. We find that in the presence of slightly deleterious mutations, MK estimates of $\alpha$ severely underestimate the true rate of adaptation even if all polymorphisms with population frequencies under 50\% are excluded. Furthermore, already under intermediate rates of adaptation, genetic draft substantially distorts the site frequency spectra at neutral and functional sites from the expectations under mutation-selection-drift balance. MK-type approaches that first infer demography from synonymous sites and then use the inferred demography to correct the estimation of $\alpha$ obtain almost the correct $\alpha$ in our simulations. However, these approaches typically infer a severe past population expansion although there was no such expansion in the simulations, casting doubt on the accuracy of methods that infer demography from synonymous polymorphism data. We suggest a simple asymptotic extension of the MK test that should yield accurate estimates of $\alpha$ even in the presence of linkage effects. 

\newpage

\section*{Introduction}

The relative importance of natural selection and random genetic drift in shaping molecular evolution is a matter of a long-standing dispute. While the neo-Darwinian synthesis placed natural selection as the dominant force~\cite{Lewontin1974}, from the late 1960's on it became popular to assume that the bulk of molecular variation is selectively neutral or at most weakly selected ~\cite{Kimura1983}. The ``neutral theory'' of molecular evolution provided an elegant explanation for the maintenance of genetic variation and the apparent constancy of the rate of molecular evolution. In addition, it enabled development of analytical approaches, based on the diffusion approximation~\cite{Kimura1964,Ewens2004}, for calculating the expected frequency spectra and fixation probabilities of polymorphisms of varying selective effect. Most commonly used approaches for estimating fitness effects of polymorphisms and demographic history rest upon these results.

Recent studies of population genomic data have strongly challenged key assumption of the neutral theory. First, in many species the rate of adaptation appears to be very high with, for example, in \emph{D.melanogaster} more than 50\% of the amino-acid changing substitutions, and similarly large proportions of noncoding substitutions, driven to fixation by positive selection~\cite{Fay2002,Smith2002,Andolfatto2005,Sella2009}. Importantly, it appears that recurrent adaptation strongly affects the genome-wide patterns of polymorphism~\cite{Andolfatto2007,Macpherson2007,Sella2009,Sattah2011}. These results imply that the dynamics of a given polymorphism is not only affected by genetic drift and purifying selection acting at its particular site, but also by the so-called genetic draft~\cite{Gillespie2000}, which describes the stochastic effects generated by recurrent selective sweeps at closely linked sites. Second, there is accumulating evidence that many polymorphisms in natural populations are slightly deleterious~\cite{Bustamante2005,Eyre-Walker2007,Lohmueller2011,Chun2011}, and such polymorphisms are expected to generate another kind of interference among linked sites, known as background selection~\cite{Charlesworth1993,Charlesworth2012}. 

It is becoming increasingly clear that the assumption of independence between sites is violated to a substantial degree in most cases in one way or another. What we do not yet fully understand is the extent to which these violations affect population genetic methods and the conclusions about the parameters of the evolutionary process inferred from such methods. It is entirely possible that their apparently wrong assumptions have only a marginal effect on the ultimate estimation. It is also possible that the estimates might be very strongly biased and generally unreliable.

Here, we focus on the investigation of one of the primary methods to test the neutral theory and to estimate the rate of adaptation at the molecular level, introduced by McDonald and Kreitman in 1991~\cite{McDonald1991}. The McDonald-Kreitman (MK) test contrasts levels of polymorphism and divergence at neutral and functional sites and uses this contrast to estimate the fraction of substitutions at the functional sites that were driven to fixation by positive selection. The MK test has been applied in many organisms with estimates of the rate of adaptation varying from extremely high in \emph{Drosophila}~\cite{Fay2002,Smith2002,Andolfatto2005, Sella2009} and \emph{E.coli}~\cite{Charlesworth2006}, to virtually zero in yeast~\cite{Elyashiv2010} and humans~\cite{Bustamante2005,Zhang2005}. These differences might reflect true variation in the rate of adaptation in different lineages or indicate that the test is biased to different extent, and possibly in different direction, in those lineages~\cite{Fay2011}.

By using closely interdigitated sites, the MK test is rather robust to many sources of error, such as variation of mutation rate across the genome and variation in coalescent histories at different genomic locations. It can be confounded, however, by slightly deleterious mutations and demography~\cite{Eyre-Walker2006b,Fay2011}. Much work has thus gone into the development of sophisticated extensions of the MK test that use the frequency distribution of polymorphisms to estimate the demographic history of the organism in question, to assess the distribution of deleterious effects at the functional sites, and to correct for both in estimating the rate of adaptation~\cite{Bustamante2002,Bustamante2005,Eyre-Walker2006a,Keightley2007,Boyko2008,Andolfatto2008,Charlesworth2008,Eyre-Walker2009,Elyashiv2010,Wilson2011}. Yet all of these extensions are still based on the assumption that evolutionary dynamics at different sites can be modeled independently of each other. In the light of the recent findings that genetic draft and background selection might often be important, it is essential to verify that these methods are robust to the linkage effects from advantageous and weakly deleterious polymorphisms and their interactions.

Unfortunately neither the current analytical nor numerical approaches based on efficient coalescent simulations are capable of modeling the interactions among multiple linked selected sites. We therefore employ large-scale forward simulations to study the effects of genetic draft and background selection on the genomic patterns of variation. This framework allows us to incorporate arbitrary distributions of fitness effects of new mutations and thus to analyze scenarios with different rates of adaptation, different strengths of positive selection, and varying levels of background selection. We use our simulations to evaluate the consistency and biases of the MK test and its extensions in re-inferring the evolutionary parameters of our simulations from the observed population diversity and divergence data. Our results reveal that the current approaches for the estimation of the rate of adaptation based on the MK test are often severely biased. We discuss various approaches for correcting these errors and highlight remaining challenges. We argue that population genetic methods that ignore linkage effects should often be misleading when adaptation is frequent.

\section*{Results}

The MK test compares the levels of diversity at neutral and potentially functional sites with the levels of divergence to evaluate whether neutral evolution can be rejected at the functional sites~\cite{McDonald1991}. An extension of the MK test can be used to estimate the fraction of substitutions driven to fixation by positive selection at the functional sites~\cite{Eyre-Walker2006b,Fay2011}: Consider the expected substitution rate at a neutral site, $d_0=2N\mu\pi_0$, where $\pi_0$ denotes the fixation probability of a neutral mutation (although $\pi_0=1/2N$, the notation of $\pi_0$ will be instructive). The rate of adaptive substitutions at a functional site, where new mutations may have arbitrary selection coefficients $s$, can be written as the difference between the overall substitution rate, minus the rate of non-adaptive substitutions:  
\begin{equation}
d_+ = d - 2N\mu \overline{\pi}= d - d_0\frac{\overline{\pi}}{\pi_0}.
\end{equation}
Here $\overline{\pi}$ specifies the average fixation probability of a non-adaptive ($s\le 0$) mutation at the functional site. The fraction of adaptive substitutions is therefore
\begin{equation}
\alpha = \frac{d_+}{d} = 1 - \frac{d_0}{d}\frac{\overline{\pi}}{\pi_0}.
\end{equation}

The ratio $d_0/d$ can be inferred from sequence alignments in the neutral and functional regions, yet estimating the ratio $\overline{\pi}/\pi_0$ is typically not straightforward. One commonly used approach is to assume that most mutations in functional regions are either neutral or highly deleterious and thus restricted to very low population frequencies~\cite{McDonald1991}, while beneficial mutations are rare and fix quickly. The observed polymorphism in the functional regions will then primarily reflect the neutral proportion of the mutation spectrum. Under this assumption, the ratio $\overline{\pi}/\pi_0$ can be approximated by the ratio $p/p_0$ between the levels of polymorphism per site in the test and the neutral reference region, yielding:
\begin{equation}
\alpha \approx 1 - \frac{d_0}{d}\frac{p}{p_0}.
\label{eq:mk_alpha}
\end{equation}

A known problem of this approach are slightly deleterious mutations. While these mutations are still unlikely to become fixed in the population, they could, however, contribute noticeably to $p$, thereby biasing estimates of $\alpha$ downwards. To minimize this problem, it has been proposed to exclude polymorphisms that are below a certain cut-off frequency~\cite{Fay2001,Charlesworth2008}; the higher this cut-off, the lower the proportion of slightly deleterious polymorphisms in the sample. More sophisticated extensions of the MK test attempt to infer the actual distribution of fitness effects (DFE) of new mutations at functional sites from the site frequency spectrum (SFS) of polymorphisms at those sites, and then correct the estimates of $\alpha$ accordingly. 

To study the effects of linkage and selection on MK-type approaches for inferring the rate of adaptation we conducted forward simulations of a 10 Mb-long chromosome with realistic gene structure, evolving under mutation, recombination, and selection over the course of $10^6$ generations. Over different simulation runs we systematically varied the rate and strength of positive selection, as well as the strength of purifying selection.

The simulated chromosome resembles a moderately gene-rich region of the human genome with approximately 4\% of its sites assumed to be functional (Materials and Methods). Note that functional density varies strongly across eukaryotes, from a few percent of constrained sites in humans to upward of 50\% in Drosophila, and the effects of linked selection should become more pronounced with higher functional density. Thus, if we find strong linkage effects in our scenario with only 4\% functional density, we would then expect even stronger effects in the functionally denser genomes such as those found in flies. In this way, our scenario should be conservative for many eukaryotic species. 

Mutations occurring at functional sites had their selection coefficients ($s$) drawn from a specified DFE, while every fourth site in exons represented a neutral, synonymous site. We assumed a mutation rate of $\mu=2.5\times 10^{-8}$ per site and generation, a recombination rate of $r=10^{-8}$ (corresponding to 1 cM/Mb), and a panmictic population of size $N=10^4$~\cite{Nachman2000,McVean2004}. These parameters are compatible with standard estimates for human evolution, such as heterozygosity at synonymous sites: $H_s=4N\mu=0.001$. Note, however, that rather than the absolute values of $\mu$, $r$, $N$, and $s$, primarily the products $N\mu$ (specifying the overall rate at which new mutations arise in the population), $Ns$ (specifying the effective strength of selection), and the ratio $s/r$ (determining the region over which a selective sweep affects the genome) should matter in our analysis. We further required that the ratio of the substitution rate at functional sites versus synonymous sites be $d/d_0\approx 0.25$, the value found in humans~\cite{Lander2001} and similar to that of many other species. This condition sets bounds on the amount of purifying selection at functional sites. 

The key observables in MK-type approaches are the levels of polymorphism and divergence at neutral and functional sites. Some approaches additionally take the SFS of polymorphism into account. In our simulations, we estimated divergence from the mutations that became fixed throughout a simulation run. Polymorphism levels and frequency distributions were estimated from population samples of 100 randomly drawn chromosomes, taken every $N$ generations throughout a run. The spectra were then averaged over all 100 samples obtained during each run. Since our chromosome has 375 kb of functional and 125 kb of synonymous sites, this corresponds to a single sample with 37.5 Mb of functional and 12.5 Mb of synonymous sites, assuming independence between samples.

In the following sections we study, in order, the effects of linkage and selection on the level of neutral polymorphism, the actual SFS of neutral and functional polymorphisms, and the fixation probabilities of deleterious mutations. At the end we analyze how all of these effects in combination affect the behavior of the MK test and its extensions.

\subsubsection*{Linkage effects on levels of neutral polymorphism}

It is well known that genetic draft and background selection reduce the levels of polymorphism at linked neutral sites~\cite{Stephan2010,Charlesworth2012}. Specifically, when strongly deleterious mutations occur at a rate $\mu_d$ per site, background selection should reduce neutral heterozygosity $H_0$ by a factor $\approx\exp(-2\mu_d/r)$~\cite{Hudson1995,Stephan1999}. Similarly, recurrent selective sweeps with selection coefficient $s_b$ occurring at rate $\nu$ per site should reduce $H_0$ by a factor $\approx(1+8K(N)\nu s_b/r)^{-1}$, where $K(N)$ is a constant~\cite{Wiehe1993,Macpherson2007}. Under a Wright-Fisher model in a diploid population of size $N$ and free recombination, we expect: $H_0=4N\mu_0$. Linkage effects from recurrent selective sweeps and background selection should then reduce $H_0$ to:

\begin{equation}
H_0 \approx 4N\mu_0 \times \frac{e^{-2\mu_d/r}}{1+8K(N)\nu s_b/r}.
\label{eq:heterozygosity}
\end{equation}

To assess the accuracy of~\EQ{heterozygosity} we compared the level of heterozygosity $H_s$ at synonymous sites in our simulation with the predicted values. Functional mutations were of four types in our simulations: neutral, beneficial, deleterious, and strongly deleterious. Each type had a specific selection coefficient: $s_n=0$, $s_d$, $s_b$, and $s_l$, respectively. We assumed that 40\% of functional mutations are always strongly deleterious~\cite{Eyre-Walker2006a,Boyko2008} and we set $s_l=-0.1$. As free parameters we chose $s_b$, $s_d$, and $\alpha$, which allowed us to assess how different strengths of purifying selection (by varying the value of $s_d$), positive selection (by varying $s_b$), and rate of adaptation (by varying $\alpha$) affect our results. Values of $\alpha$ in our simulations ranged from 0 to 0.5, $s_b$ from 0.001 to 0.05, and $Ns_d$ from -1 to -100 (Table S1).

Fig.~1A shows that inferred and predicted levels of neutral heterozygosity are generally in good agreement. Only when adaptation was very frequent and strong the predicted reduction in $H_s$ is slightly overestimated. The amount by which linkage effects reduce $H_s$ is primarily determined by the product of rate and strength of adaptation (Fig.~1A, inset). The contribution of background selection is typically less severe and appears most pronounced for the very weakly deleterious selection coefficients, as indicated by the observation that for the same value of $\alpha s_b$, the simulation runs with the weaker deleterious selection coefficients ($Ns_d\approx -1$, darker points in the inset) yield stronger reduction of $H_s$.

\subsubsection*{Linkage effects on the SFS at functional and synonymous sites}

In applications of the MK test and its extensions it is often not only the level of polymorphism that is important, but also the SFS at functional and neutral sites. Some heuristic methods simply eliminate low-frequency variants, while some, more sophisticated, methods try to infer the actual DFE at functional sites from the SFS.

In the Wright-Fisher model under mutation-selection-drift balance and free recombination, the average number of polymorphism where the derived allele is present at population frequency $x$ is given by~\cite{Wright1938,Sawyer1992}
\begin{equation}
g(x,s) = 4N\mu_s \frac{1-e^{-4Ns(1-x)}}{(1-x)x(1-e^{-4Ns})}.
\label{eq:ms_balance}
\end{equation} 
Here $\mu_s$ is the rate at which new mutations with selection coefficient $s$ arise at the locus of interest per generation per individual. Integrated over the full DFE of new mutations, as specified by a density function $\rho(s)$, the expected SFS for all polymorphism at the locus is then $g(x)=\int g(x,s)\rho(s)ds$. 

It is well know that genetic draft and background selection can distort the SFS from this expectation~\cite{MaynardSmith1974,Wiehe1993,Braverman1995,Fay2000,McVean2000,Bustamante2001,Lohmueller2011,Sattah2011}. What is not clear is whether the deviations are marked under realistic evolutionary scenarios and whether this might affect population genetic methods based on the assumption of mutation-selection-drift balance. We measured the SFS at functional and synonymous sites in our simulations and compared it with the prediction under mutation-selection-drift balance given the DFE of the particular simulation run. In an attempt to account for the reduction in overall levels of diversity and reduced effectiveness of selection due to genetic draft and background selection, we replaced $N$ in~\EQ{ms_balance} by the effective population size inferred from the level of heterozygosity at synonymous sites, $H_s=4N_e\mu_0$, in the particular simulation run.  

The left plot in Fig.~1B shows the observed and expected SFS at functional and synonymous sites in our simulations for a scenario with no adaptation but high levels of background selection ($Ns_d=-2$). Expected and observed spectra are in good agreement, suggesting that for the chosen recombination rate ($r=10^{-8}$) and functional density ($\approx4\%$ of the chromosome) the effects of background selection alone are well approximated by mutation-selection-drift balance with $N_e$ being adjusted to the value obtained from the level of heterozygosity at neutral sites. This shows that in the presence of background selection alone it should be possible to estimate the DFE at functional sites reasonably well.

However, deviations between observed and expected spectra become noticeable once adaptation becomes more frequent (Fig.~1B, middle plot). The right plot in Fig.~1B shows a scenario with frequent adaptation ($\alpha=0.5$) and strong sweeps ($s_b=0.05$). Here the deviations between observed and expected spectra are substantial at both synonymous and functional sites. Intermediate frequency polymorphisms are depleted while there is an excess at high and low derived allele frequencies compared to the expectation under mutation-selection-drift balance at both functional and neutral sites. These distortions do not fit any model of mutation-selection-drift balance with a constant effective population size, suggesting that methods that use such models to infer the DFE from the SFS at functional sites might run into severe biases in the presence of even moderate levels of adaptation.

\subsubsection*{Linkage effects on fixation probabilities of deleterious mutations}

Levels of divergence at functional and neutral sites are the other key parameters that are used in the MK test and its extensions. Linked selection cannot affect the rate of neutral divergence as it is always equal to the rate of mutation at neutral sites. The rate of divergence at functional sites, however, could be affected substantially.

In the Wright-Fisher model under free recombination, a mutation with selection coefficient $s$ that arises in one individual of a diploid population of size $N$ eventually fixes with probability:

\begin{equation}
\pi(s) = \frac{1-e^{-2s}}{1-e^{-4Ns}}.
\label{eq:p_fixation}
\end{equation}

Genetic draft and background selection are expected to increase the fixation probabilities of deleterious mutations: Under recurrent selective sweeps, deleterious mutations can hitchhike to frequencies they are unlikely to reach under mutation-selection-drift balance alone, increasing their chance of fixation over that expected without linkage~\cite{Chun2011,Hartfield2011}. Similarly, background selection renders purifying selection less effective by reducing the number of successfully reproducing individuals, thereby also increasing the fixation probabilities of deleterious mutations~\cite{Barton1995,Hartfield2011,Charlesworth2012}. 

One common approach for addressing these issues is to assume that~\EQ{p_fixation} can still be used but that~$N$ has to be replaced by a lower, effective population size~$N_e$. However, it is not clear whether a single scalar~$N_e$ applies over a range of selection coefficients. We tested this in our simulations by measuring the fixation probabilities of deleterious mutations with different selection coefficients $s_b$ and then inferring the corresponding values of $N_e$ according to~\EQ{p_fixation} for the different selection coefficients in the same run independently. Every simulation run had a particular rate ($\alpha$) and strength ($s_b$) of adaptation, while deleterious functional mutations had selection coefficients $s_b=-0.001$, $-0.0005$, $-0.0002$, and $-0.0001$, with all four classes being of equal proportion. The fraction of neutral mutations at functional sites was again tuned to yield $d/d_0\approx 0.25$.

Fig.~1B shows the inferred values of $N_e$ according to~\EQ{p_fixation} as a function of $s_d$. Our results confirm that genetic draft and background selection generally increase fixation probabilities of deleterious mutations, as indicated by the fact that the inferred~$N_e$ is always smaller than the actual $N=10^4$. However, in the same simulation run different selection coefficients have very different values of inferred $N_e$. For example, in the simulation run with $s_b=0.001$ and $\alpha=0.17$, the mutations with $s_d=0.0001$ fix with a probability that corresponds to $N_e\approx 8500$, while the mutations with $s_d=0.001$ yield $N_e\approx 5500$. For stronger sweeps and higher $\alpha$ the discrepancies become even more profound. In none of the investigated scenarios we found a scalar $N_e$ that works for all four deleterious selection coefficients. 

Note that because $N$ enters~\EQ{p_fixation} exponentially, differences in $N$ yield substantial differences in the actual fixation probabilities. In the above scenario, for instance, the 30\% difference between $N_e\approx 5500$ and $N_e\approx 8500$ corresponds to an approximately 400-fold difference in the fixation probability for mutations with $s_d=-0.0005$. Note also that $N_e$ according to the fixation probabilities of deleterious mutations is typically much lower than $N_e$ inferred from the levels of neutral heterozygosity according to $H_s=4N_e\mu$, except for very weakly deleterious mutations.

These results indicate that there is no scalar transformation of $N_e$ that would allow us to estimate fixation probabilities across multiple fitness classes. Thus, even if we were to know the true DFE at functional sites, it would still be impossible to use mutation-selection-drift methods to predict the rate of fixation of deleterious mutations under scenarios that include even moderate amounts of genetic draft.

\subsubsection*{MK estimates of the rate of adaptation}

In the previous sections we have shown that linked selection can affect the key quantities in the MK test in complex ways that do not fit the predictions under mutation-selection-drift balance. However, some of the errors partially compensate for each other in the context of the MK test. For example, genetic draft might cause deleterious mutations to appear virtually neutral in the polymorphism data (they could be present at unexpectedly high frequencies) but would also elevate their probabilities of fixation to that of neutral mutations. It is thus possible that the effects we described above might generally not affect MK estimates of $\alpha$ strongly.

Our simulations allow us to explicitly test the accuracy of MK estimates of $\alpha$ inferred from~\EQ{mk_alpha}. Fig.~2 shows the comparison of true values and MK estimates for all simulation runs from Table S1. Polymorphism levels $p$ and $p_0$ were again calculated from samples of $100$ genomes drawn every $N$ generations; substitution rates $d$ and $d_0$ were inferred from the mutations that became fixed over the course of a simulation run. To minimize the bias generated by slightly deleterious polymorphisms, we considered only polymorphisms with a derived allele frequency of $x\ge 0.1$ (Figure 2, left panel) or $x\ge 0.5$ (Figure 2, right panel) in the samples. Our results demonstrate that MK estimates of $\alpha$ under both cut-offs still tend to underestimate $\alpha$, often substantially. For example, when the true $\alpha=0.4$, the MK estimate using a cut-off $x\ge 0.1$ yields a negative value of $-0.2$ for a scenario where $s_b=0.001$ and $Ns_d=-1$. Increasing the cut-off from $x\ge 0.1$ to $x\ge 0.5$ reduces this discrepancy, but substantial errors remain. In the above scenario with $\alpha\approx 0.4$ the MK estimate still yields only $\alpha\approx 0.18$. 

The underestimation of $\alpha$ is generally more pronounced when deleterious mutations are only weakly deleterious than when they are strongly deleterious. This is consistent with weakly deleterious mutations having a higher chance of contributing to polymorphism than strongly deleterious mutations, but still having low probabilities of fixation, thus yielding higher overestimates for $\overline{\pi}/\pi_0$ based on $p/p_0$. Strongly deleterious mutations contribute to neither polymorphism nor divergence and thus do no bias estimates of $\alpha$. As strength of positive selection increases, the biases due to weakly deleterious mutations can be mitigated to some extent because now they become effectively neutral and contribute to both polymorphism and divergence.

\subsubsection*{DFE-based extensions of the MK approach}

Several methods for correcting possible biases in MK estimates have been proposed that go beyond the simple exclusion of low-frequency polymorphisms. These methods aim to first estimate the DFE at functional sites and then calculate how many non-adaptive mutations are expected to become fixed given the inferred DFE~\cite{Bustamante2002,Bustamante2005,Eyre-Walker2006a,Keightley2007,Eyre-Walker2007,Boyko2008,Eyre-Walker2009}. Any excess of substitutions should be attributable to adaptation. Some approaches additionally aim to correct for possible effects of demography, which is first inferred from the SFS at synonymous sites and then used for correcting the SFS at functional sites~\cite{Williamson2005,Keightley2007,Boyko2008}.     

One particularly popular such method is DFE-alpha by Eyre-Walker and Keightley~\cite{Eyre-Walker2009}. Here we investigate the performance of this method as a representative of the class of methods based on the same paradigm~\cite{Bustamante2002,Bustamante2005,Eyre-Walker2006a,Keightley2007,Eyre-Walker2007,Boyko2008,Eyre-Walker2009}. DFE-alpha models the DFE at functional sites by a gamma distribution, specified by the mean strength of selection, $\gamma=-N_e \overline{s}$, and a shape parameter $\beta$, allowing the distribution to take on a variety of shapes ranging from leptokurtic to platykurtic. DFE-alpha incorporates two simple demographic models: (i) constant population size and (ii) a single, instantaneous change in population size from an ancestral size $N_1$ to a present-day size $N_2$ having occurred $t$ generations ago. Provided the SFS at both neutral and functional sites and the respective levels of divergence, DFE-alpha infers $\gamma,\beta,N_2/N_1, t$, and $\alpha$ at functional sites.

We applied DFE-alpha to polymorphism and divergence data from our simulations (Materials and Methods). For this analysis, we modified our simulations such that the selection coefficients of the non-adaptive mutations at functional sites were drawn from a gamma-distribution and thus the same distribution was used in the simulations that is assumed by DFE-alpha. We chose a shape parameter of $\beta=0.2$, resembling empirical estimates from polymorphism data at non-synonymous sites in humans~\cite{Keightley2007,Eyre-Walker2007,Boyko2008}. We varied $\alpha$ from 0 to 0.5 and investigated two scenarios with $s_b=0.001$ or $s_b=0.01$. The mean of the DFE was tuned for each scenario such that $d/d_0\approx 0.25$. Throughout our simulations population size was always kept constant at $N=10^4$ individuals.

Table 1 shows the performance of DFE-alpha under its two demographic models. When using the correct model of constant population size, DFE-alpha systematically overestimates $\alpha$ and underestimates the strength of selection against deleterious mutations. The shape parameter $\beta$ of the gamma distribution is overestimated by almost two-fold under strong and frequent adaptation. These biases are generally more pronounced for the scenarios with stronger sweeps than for those with weaker sweeps. Under the model with a population size change, the estimates of $\alpha$ and $\beta$ become more accurate (within $\pm0.05$ of true values) but the mean strength of selection against deleterious mutations is now overestimated by roughly 50\%. Strikingly, under this model DFE-alpha always infers a substantial population size expansion while there was no such expansion in our simulation.

This behavior of DFE-alpha is consistent with the fact that genetic draft leaves signatures in the SFS similar to those observed under a recent population size expansion, namely a skew towards low-frequency polymorphisms. The extent of this effect, however, is alarming, given that even for a scenario where $\alpha$ is only about 0.1 already an almost 10-fold population size expansion is inferred by DFE-alpha (which is a built-in limit of DFE-alpha as currently implemented). Note that even in the scenario with no adaptation DFE-alpha still infers a 5-fold population size expansion, implying that background selection alone can already bias demographic inference.

Thus, it appears that methods such as DFE-alpha, where a demographic model is first fit to the SFS at synonymous sites, indeed infer reasonable estimates of $\alpha$ while entirely misinterpreting demography and also overestimating the strength of purifying selection. The reason seems to be that the correction for demography these approach attempt to provide, in our scenario with a constant population size, instead serves as a correction for the effects of genetic draft on the SFS. This correction can work well for the estimation of $\alpha$ but not for the estimates of the strength of purifying selection.

\section*{Discussion}

It is well known that linkage effects among loci, such as genetic draft and background selection, can affect the patterns and dynamics of molecular variation~\cite{Hill1966,MaynardSmith1974,Wiehe1993,Charlesworth1993,Braverman1995,Fay2000,Barton1995,Hartfield2011,Chun2011,Charlesworth2012}. In this study, we have used forward simulations that explicitly incorporate linkage and selection on a chromosome-wide scale to investigate quantitatively how linked selection biases common population genetics methods. We specifically tested the performance of the MK test and its extensions to infer the rate of adaptation. 

Consistently with previous results~\cite{Charlesworth2008}, we found that MK estimates of the rate of adaptation can be severely biased in the presence of slightly deleterious mutations and generally underestimate $\alpha$. Unfortunately our analysis shows that the standard approaches to address this known problem do not typically resolve it: 
 
(i) The simple heuristic approach, where low-frequency polymorphisms are excluded from the analysis, renders MK estimates more accurate, but a substantial bias remains (Fig.~2). The reason for this is that the dynamics of slightly deleterious polymorphisms under recurrent selective sweeps can be very different from the expectation under the diffusion model, which predicts that frequent mutations should have a realistic chance of eventually reaching fixation. However, under recurrent selective sweeps a slightly deleterious mutation can easily hitchhike to substantial population frequencies yet become unlinked during the late phase of a sweep. This deleterious mutation can then spend substantial time as a frequent polymorphism in the population while it slowly declines in frequency. At every stage of this process, the frequency of the mutation substantially overestimates its fixation probability. Such mutations are not effectively removed from a population sample by excluding low-frequency polymorphisms. 

(ii) Modern extensions of the MK test aim to address the problem of slightly deleterious mutations by estimating the actual contributions of deleterious mutations to polymorphism and divergence. We found that these approaches misestimate the mean and the shape of the DFE and, as a result, tend to overestimate the rate of adaptation (Table 1). This is not surprising given that such approaches infer the DFE at functional sites by fitting the observed SFS to that predicted under mutation-selection-drift balance, which can be substantially distorted by linkage effects (Fig.~1C).           

(iii) The most sophisticated extensions of the MK test available today additionally attempt to correct for demography. These approaches try to infer demographic history from the SFS at putatively neutral (typically synonymous) sites, and this demography is then incorporated into the estimation procedure for the DFE at functional sites. Interestingly, we found that such methods obtain accurate estimates of the rate of adaptation while inferring erroneous demography and also inaccurate estimates of the mean strength of purifying selection against functional mutations (Table 1). This seeming contradiction reflects the fact that the distortions of the SFS at synonymous sites, which these methods interpret to be due to demography, can in fact be due to genetic draft. As we have shown in Fig.~1C, these distortions are very similar at synonymous and functional sites. Thus, by imposing a demographic scenario that corrects for distortions of the SFS at synonymous sites, the methods also correct the SFS at functional sites. 

The fact that methods such as DFE-alpha seem to obtain accurate estimates of $\alpha$ under a ``demographic correction'', suggests that a simple heuristic extension of the standard MK test, where the effects of genetic draft on the SFS at synonymous and functional sites are simply divided out, might already provide reasonable estimates without having to invoke demography. To illustrate such an approach, let us define $\alpha(x)$ as a function of the frequency of the derived mutations:

\begin{equation}
\alpha(x) = 1 - \frac{d_0}{d}\frac{p(x)}{p_0(x)}.
\label{eq:as_mk}
\end{equation} 
Here $p(x)$ and $p_0(x)$ are the numbers of polymorphism at functional and synonymous sites, respectively, with derived allele at frequency $x$. Because $\alpha(x)$ depends only on the ratio $p(x)/p_0(x)$, any biases affecting the SFS at functional and synonymous sites in the same way, regardless whether due to demography or genetic draft, effectively cancel out. Furthermore, we can extrapolate $\alpha(x)$ to $x\to 1$, where it should asymptotically converge to the true $\alpha$, assuming that adaptive mutations do not significantly contribute to polymorphism and that purifying selection has been sufficiently stable over time.

As a proof of principle, we show in Fig.~3A that this simple heuristic extension of the MK approach indeed converges asymptotically to the true value of $\alpha$ in our simulations, even in a scenario with a high rate of adaptation ($\alpha=0.42$), strong sweeps ($s_b=0.01$), and slightly deleterious mutations ($Ns_d=-2$). While it is not straightforward to predict the precise functional form of $\alpha(x)$, which will depend on the specific DFE, fitting an exponential approximation of the form $\alpha(x)\approx a+b\exp{(-cx)}$ seems to work reasonably well. Fig.~S1 shows the comparison between asymptotic MK estimates obtained by this procedure and the true values of $\alpha$ for all simulation runs from Supplementary Table S1. In Fig.~3B we compare the true values of $\alpha$ for all simulation runs from Table 1, the respective standard MK estimates using a cut-off of $x\ge 0.1$, and the estimates from DFE-alpha under its two demographic models. The asymptotic MK estimates no longer suffer from a systematic downward bias due to deleterious mutations and are much more accurate than standard MK estimates, as well as estimates from DFE-alpha without the ``demographic correction''. They are similarly accurate to estimates from DFE-alpha with the correction.  

In order to further verify that this simple asymptotic MK approach yields similar results as the more complex approaches invoking ``demographic correction'' and estimation of the DFE, we applied asymptotic MK to previously analyzed polymorphism and divergence data from \emph{D.~melanogaster} and humans (Fig.~3C). The human data consists of $11,000$ protein-coding regions that had been resequenced by Celera Genomics in 20 European American individuals~\cite{Bustamante2005}. After excluding polymorphisms with frequencies below 10\% or above 90\%, we obtained an asymptotic MK estimate of $\alpha=0.13$ $(0.09,0.19)$ for this data. This is consistent with the range of $\alpha=0.1-0.2$ estimated in~\cite{Boyko2008}. Note that the standard MK estimate for this data when excluding all polymorphisms with sample-frequencies below 10\% yields a negative value $\alpha=-0.05$. For \emph{D. melanogaster}, we obtained an estimate of $\alpha=0.57$ $(0.54,0.60)$ using polymorphism data from 162 inbred lines derived from Raleigh, North Carolina by the \emph{D.melanogaster} Genetic Reference Panel~\cite{Mackay2012}. This estimate is similar, although somewhat higher, than previously estimated values obtained from earlier polymorphism data sets in this species~\cite{Fay2002,Smith2002,Andolfatto2005,Sella2009,Mackay2012}. 

The results presented in this study have important ramifications for the inference of evolutionary parameters from polymorphism and divergence data. It appears that despite the complexity of the process, we do have means of estimating the rate of adaptive evolution by using DFE-alpha like approaches with the ``demographic correction'' or use the simple asymptotic MK test we suggested above. It is important to consider that the standard MK approach with or without excluding rare polymorphisms produces severely biased estimates under many scenarios and even when adaptation is not pervasive. 

Unfortunately, estimation of the DFE and, especially, of demography tend to be severely affected by already moderate amounts of genetic draft and background selection. Estimating demography from neutral sites that are close to functional ones (such as synonymous sites) should in general lead to erroneous inference of population expansions. One solution would be to use regions that have very low functional density and a high recombination rate for such inference. It remains to be determined which genomic regions are appropriate in this way, for example in the human genome.

Our analysis suggest that in the presence of genetic draft and background selection the evolutionary interactions among linked polymorphisms of different selective effects are complex and consequential. It is clear that the standard diffusion approximation that attempts to model evolution at different sites independently and wrap the complexity of linkage effects among sites into effective parameters such as~$N_e$, can introduce massive errors into the estimation of key population genetic parameters. We thus believe that new analytics need to be developed that correct for linkage effects. When diffusion fails, other approaches such as stochastic jump processes might succeed. It is also important to develop new approaches that use forward simulations under realistic scenarios of genetic draft and background selection to estimate evolutionary parameters of interest. At the very least, one has to verify with forward simulations, such as the one presented here (SLiM) or similar programs~\cite{Hernandez2008,Chadeau2008,Carvajal2008a}, that commonly used heuristic and analytic methods in population genetics are robust to linkage effects.

\section*{Materials and Methods}

\subsubsection*{Forward simulations of chromosome evolution}

Our simulations model the population dynamics of a 10 Mb-long chromosome evolving in a panmictic diploid population under mutation, recombination, and selection. Genes are placed equidistantly on the chromosome with a density of one gene per 40 kb~\cite{Lander2001}. Each gene consists of 8 exons of length 150 bp each, separated by introns of length 1.5 kb. Genes are flanked by a 550 bp-long 5' UTR and a 250 bp-long 3' UTR. We assume that three out of four sites in exons and UTRs are functional sites. Every 4th site in exons and UTRs is non-functional with all mutations at those sites being neutral. These non-functional sites are used to model synonymous sites. Mutations occurring outside of exons or UTRs are neutral. Altogether, this yields a functional fraction of~$3.75\%$ of the chromosome.

For each chromosome we store the list of mutations it harbors, with each mutation being specified by its position along the chromosome and its selection coefficient. The population consists of $N=10^4$ diploid individuals. We assume that mutations are codominant and that fitness effects at different sites in the genome are additive. The fitness of an individual is thus given by $w=1+\sum_i s_i$, where the sum is taken over the selection coefficients, $s_i$, of all mutations on its two chromosomes.

Population dynamics is simulated in a model with discrete generations and constant population size. In each generation, a set of $N=10^4$ children is newly generated. The two parents of each child are drawn from the population in the previous generation with probabilities proportional to their fitnesses. To generate the haploid gamete a parent contributes to the child, the two parental chromosomes undergo recombination at a uniform rate of $r=10^{-8}$ per site along the chromosome (corresponding to 1 cM/Mb). Each gamete then undergoes mutation, where new mutations occur at a rate $\mu=2.5\times 10^{-8}$ per site per generation uniformly along the chromosome. Only the mutations which fall into exons or UTRs are followed in our simulations.

While every mutation has a specific position along the chromosome, the simulation makes an infinite sites assumption in the sense that a chromosome can harbor more than one mutation at the same site and that back-mutations do not occur. Given our population parameter $N\mu=2.5\times 10^{-4}$, the choice of an infinite sites model is well justified. The simulation does not model the actual nucleotide states of mutations. The selection coefficient of each new mutation, if it falls at a functional site, is drawn from a specific DFE. Mutations that fall at non-functional sites always have $s=0$. After all $N=10^4$ children have been generated this way, their fitnesses are calculated and they become the parents for the next generation.

At the start of a simulation run all individuals are initialized with empty chromosomes since no mutations have yet occurred. The simulations then go through a burn-in period of $10N$ generations to establish a stationary level of diversity. Every 100 generations the population is screened for fixed mutations, i.e., mutations that are present in all individuals of the population. These mutations are recorded as substitutions and removed from all chromosomes for they can no longer cause fitness differences between individuals. A simulation run is followed for  $10^6$ generations after the burn-in. 

The simulation is implemented in C++, making extensive use of algorithms from the GNU scientific library~\cite{Galassi2009}. An extended version of the simulation is implemented in the open-source program SLiM, which can be downloaded from the author's homepage at: www.stanford.edu/$\sim$messer/software. The website also provides a comprehensive documentation for the program and several application examples. 

\subsubsection*{DFE-alpha estimation on simulation data}

We ran DFE-alpha for each of the simulation runs specified in Table 1 using the online server provided at: http://homepages.ed.ac.uk/eang33/software. These runs simulated the evolution of the above described 10 Mb-long chromosome in a population of $N=10^4$ diploid individuals over the course of $10^6$ generations under the specific selection scenario. The SFS at functional and synonymous sites were calculated from samples of 100 randomly drawn chromosomes, taken every $N$ generations in a simulation run. The SFS obtained from each sample were then averaged over all 100 samples taken throughout each run to generate the unfolded spectra provided to DFE-alpha. Since our 10 Mb-long chromosome has 375 kb of functional and 125 kb of synonymous sites, this corresponds to a single sample with 37.5 Mb of functional and 12.5 Mb of synonymous sites, assuming independence between samples. Divergence counts at functional and synonymous sites were inferred from the observed substitutions in each simulation run.   

\subsubsection*{Asymptotic MK estimation in humans and flies}

Human polymorphism and divergence data are based on the re-sequencing of 11,404 protein coding-genes in 20 European American individuals and were obtained from Table S2 in~\cite{Boyko2008}. A detailed description of the sequencing is provided in~\cite{Bustamante2005}. Polymorphism data for \emph{D.melanogaster} was obtained from the genome sequences of 162 inbred lines derived from Raleigh, North Carolina~\cite{Mackay2012}. Only coding regions with sequence information for at least 130 strains and one-to-one orthologs across the 12 Drosophila species tree~\cite{Clark2007} were considered in our analysis. Each SNP was down-sampled to 130 strains and SNPs that were no longer polymorphic after the down-sampling were removed. Divergence data with \emph{D. simulans} was obtained from PRANK alignments of the 12 Drosophila species. Ancestral SNP states were determined via parsimony to \emph{D. simulans}. Functional annotation was obtained from Flybase release 5.33~\cite{Flybase2012}.

\section*{Acknowledgments}

We thank David Lawrie for processing the \emph{D.melanogaster} polymorphism and divergence data for our asymptotic MK estimation in Figure 3C. We thank Daniel Fisher, Peter Keightley, Adam Eyre-Walker, David Enard, Nandita Garud, and members of the Petrov lab for helpful discussions and comments on the manuscript. The work was supported by the NIH under grants RO1GM100366, RO1GM097415, and R01GM089926  to DAP.

\singlespacing

{\bibliography{/Users/messer/Dropbox/bibliography/bibliography}}

\newpage

\doublespacing

\begin{table}[h!]
\centering
\begin{tabular}{cccc|ccc|cccccccc}
\multicolumn{4}{c|} {simulation values} &\multicolumn{3}{c|}{ DFE $\alpha$ (constant $N$)} & \multicolumn{5}{c}{DFE $\alpha$ (step-change)} \bigstrut\\
\hline
$s_b$ & $\alpha$ & $\gamma$ & $\beta$ & $\alpha$ & $\gamma$ & $\beta$ & $\alpha$ &  $\gamma$ & $\beta$ & $N_2/N_1$ & $t/N_2$ \bigstrut \\
\hline
- & 0.00 & 448 & 0.2 & 0.12  & 297 & 0.26 & 0.00 & 703 & 0.21 & 5.0 & 6.2\\
\hline
0.001 & 0.05 & 434 & 0.2 & 0.20 & 264 & 0.27 & 0.07 & 676 & 0.21 & 5.0 & 5.4\\
0.001 & 0.09 & 437 & 0.2 & 0.22 & 288 & 0.26 & 0.09 & 914 & 0.20 & 8.8 & 5.2\\
0.001 & 0.18 & 441 & 0.2 & 0.27 & 265 & 0.27 & 0.15 & 754 & 0.21 & 8.8 & 5.4\\
0.001 & 0.28 & 422 & 0.2 & 0.40 & 276 & 0.27 & 0.29 & 1055 & 0.20 & 10.0 & 4.6\\
0.001& 0.37 & 836 & 0.2 & 0.50 & 354 & 0.28 & 0.41 & 1250 & 0.21 & 10.0 & 4.7\\
0.001 & 0.49 & 1638 & 0.2 & 0.57 & 532 & 0.29 & 0.48 & 2438 & 0.21 & 10.0 & 4.2\\
\hline
0.01 & 0.06 & 424 & 0.2 & 0.24 & 233 & 0.27 & 0.11 & 635 & 0.21 & 5.0 & 4.9\\
0.01 & 0.09 & 424 & 0.2 & 0.26 & 217 & 0.29 & 0.12 & 675 & 0.22 & 10.0 & 4.7\\
0.01 & 0.18 & 381 & 0.2 & 0.40 & 152 & 0.31 & 0.24 & 654 & 0.21 & 10.0 & 3.5\\
0.01 & 0.27 & 339 & 0.2 & 0.49 & 109 & 0.34 & 0.31 & 618 & 0.21 & 10.0 & 2.8\\
0.01 & 0.36 & 652 & 0.2 & 0.58 & 158 & 0.35 & 0.43 & 1113 & 0.22 & 10.0 & 2.6\\
0.01 & 0.47 & 1154 & 0.2 & 0.68 & 182 & 0.38 & 0.53 & 1802 & 0.22 & 10.0 & 2.1\\
\end{tabular}
\end{table}

\noindent {\bf Table 1.} Performance of DFE-alpha under its two demographic models. Each row is a particular simulation run with the evolutionary parameters specified in the left four columns. The average strength of purifying selection, $\gamma=-4N_e\overline{s}$, was calculated from the mean of the DFE used in the simulation and $N_e$ inferred from heterozygosity at synonymous sites. The middle three columns show the estimates from DFE-alpha under the demographic model with constant population size. The last five columns show the estimates under the demographic model with a single population size change. $N_2/N_1$ is the inferred ratio between present and ancient population size, $t$ is the estimated time since the population size change.

\newpage

\begin{figure}[h!]
\centering
\includegraphics[width=\linewidth]{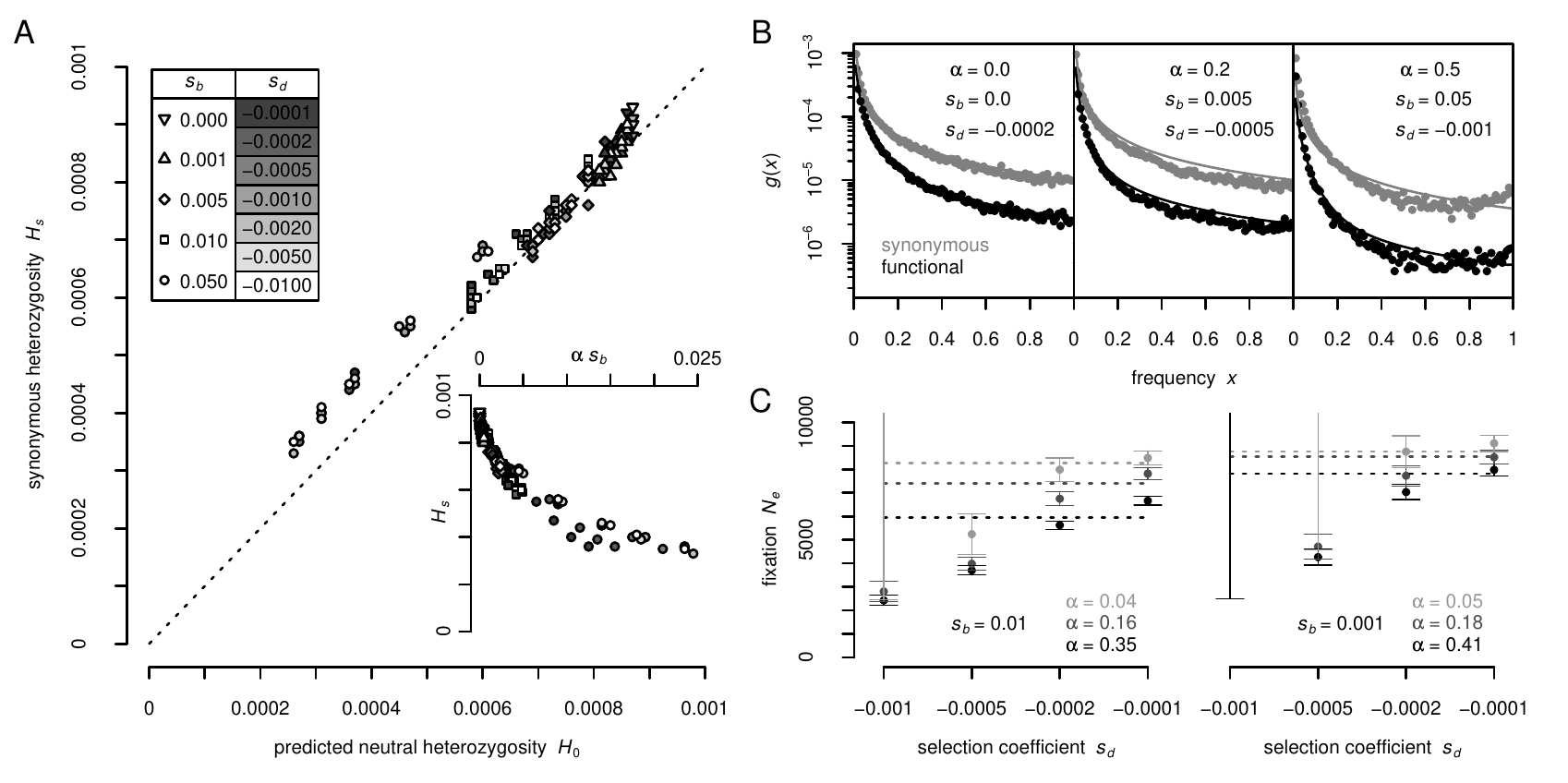}
\end{figure}

\noindent {\bf Fig.~1.} (A) Observed levels of heterozygosity at synonymous sites in our simulations compared with the predicted level according to~\EQ{heterozygosity} for each simulation run. The inset shows $H_s$ as a function of $\alpha s_b$. (B) SFS at functional and synonymous sites in three different simulation runs. Symbols show the observed numbers of polymorphisms per site averaged over all population samples taken throughout the run. Lines show the expected spectra under mutation-selection-drift balance using the value of $N_e$ inferred from heterozygosity at synonymous sites. Expected spectra were corrected for binomial sampling. The left plot shows the results for a run with no adaptation and strong background selection, the middle plot shows a scenario with an intermediate rate of adaptation, the right plot shows a scenario with frequent and strong adaptation. (C) Effective population sizes estimated from the observed fixation probabilities of deleterious mutations according to~\EQ{p_fixation}. The left plot shows three simulation runs with different rates of adaptation and $s_b=0.01$. The right plot shows three runs with weaker strength of positive selection ($s_b=0.001$). The four different deleterious selection coefficients always yield very different values of $N_e$. Dashed lines indicate the value of $N_e$ inferred from the level of synonymous heterozygosity according to $H_s=4N_e\mu_0$. Error bars are Pearson 95\% confidence intervals assuming that fixations of deleterious mutations are described by a Poisson process.

\newpage

\begin{figure}[h!]
\centering
\includegraphics[width=0.5\linewidth]{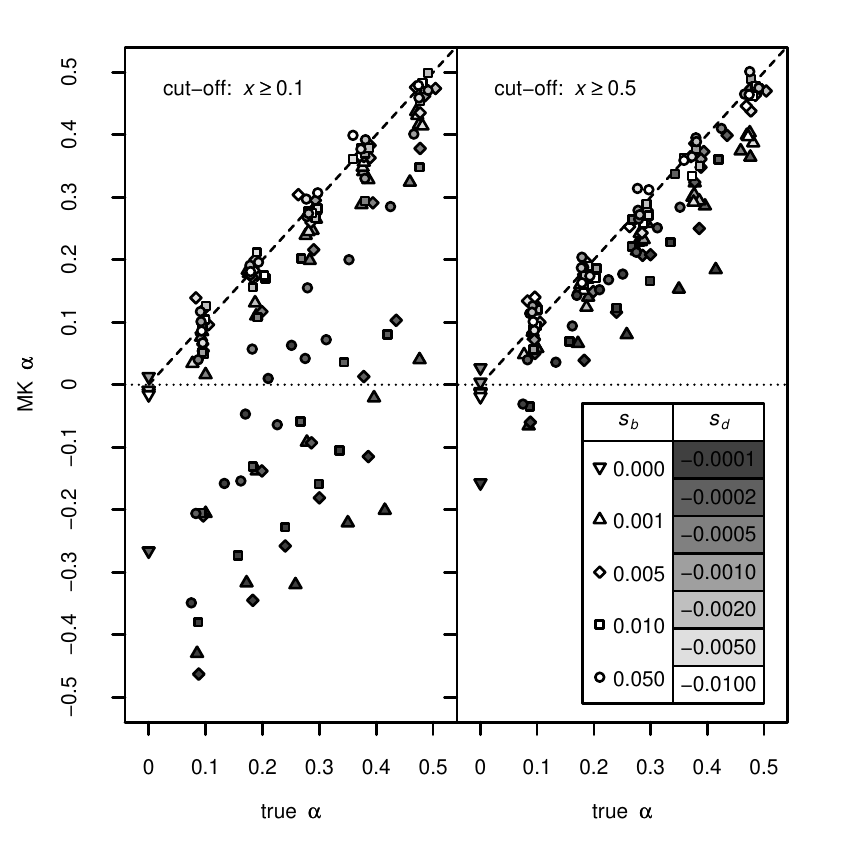}
\end{figure}

\noindent {\bf Fig.~2.} Comparison of the true values of $\alpha$ and MK estimates according to~\EQ{mk_alpha} obtained from the observed levels of polymorphism and divergence at synonymous and functional sites in all simulation runs from Table S1. On the left, results are shown for a cut-off derived allele frequency of $x\ge0.1$. On the right, results are shown for a cut-off $x\ge 0.5$.

\newpage

\begin{figure}[t!]
\centering
\includegraphics[width=\linewidth]{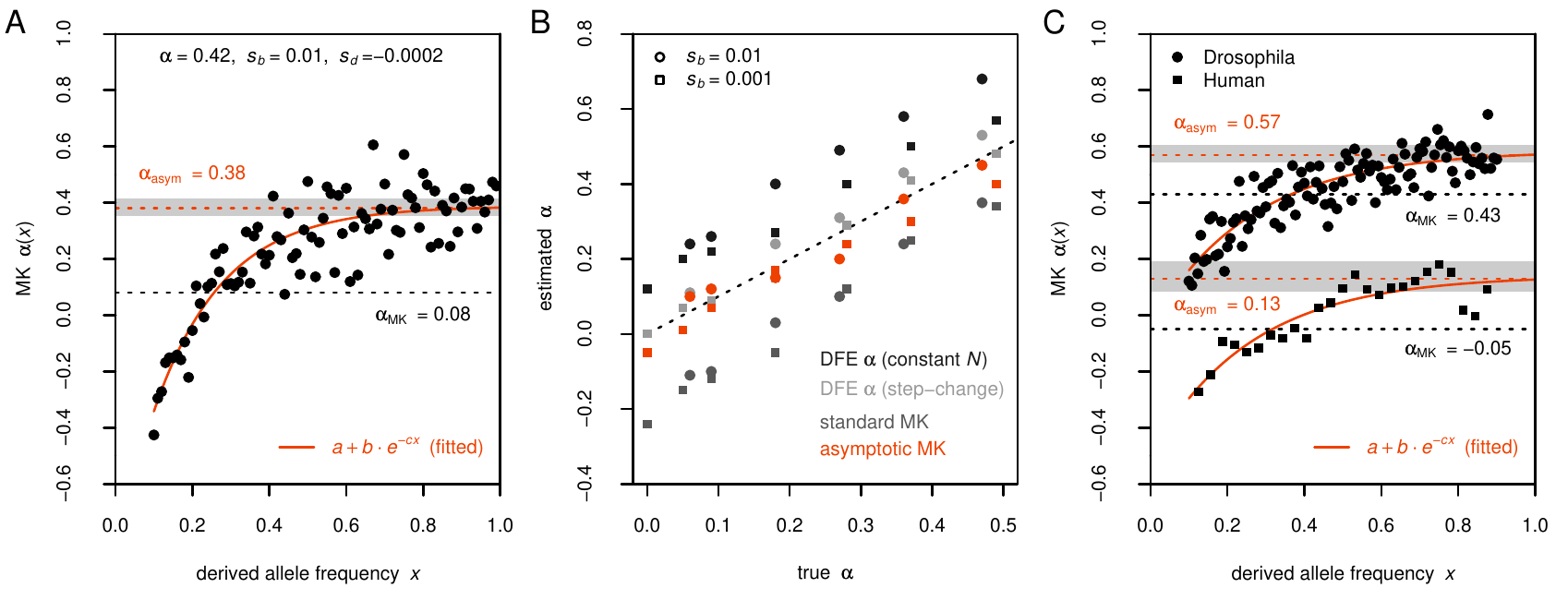}
\end{figure}

\noindent {\bf Fig.~3.} (A) Asymptotic MK estimation for a simulation run with $s_b=0.01$, $s_d=-0.0002$, and $\alpha=0.42$. The standard MK estimate using a cutoff $x\ge 0.1$ yields $\alpha=0.08$ (dashed black line). The asymptotic MK estimate yields $\alpha=0.38$ and was obtained by fitting an exponential function $\alpha(x)=a+b\exp{(-cx)}$ for all $x\ge 0.1$ using nonlinear least-squares and extrapolating to $x=1$ (dashed red line). The grey bar denotes the area between the 5\% and 95\% quantiles obtained from 1000 bootstrap samples (the observed values $\alpha(x_i)$ were resampled and the resampled sets were then fit). (B) Comparison of true values of $\alpha$ for the simulation runs from Table~1 with DFE-alpha estimates under its two demographic models, standard MK estimates using a cutoff-frequency $x\ge0.1$, and asymptotic MK estimates. Circles show data for runs with $s_b=0.01$, squares show the data for runs with $s_b=0.001$. (C) Asymptotic MK estimation at nonsynonymous sites in humans and Drosophila. The dashed black lines show the respective standard MK estimates using a cutoff $x\ge 0.1$ and grey bars again denote the areas between the 5\% and 95\% quantiles obtained from 1000 bootstrap replicates.

\newpage

\vspace{1cm}

\newpage

\subsection*{Table S1}

\singlespacing

{\small

\begin{tabular}{c|c|c|c|c|c}
$\rho_b$ & $s_b$ & $s_d$ & true $\alpha$ & MK $\alpha$ $(0.1)$ & MK $\alpha$ $(0.5)$\bigstrut \\
\hline
0.000000 & na & -0.0001 & 0.00 & -0.55 & -0.16 \\
0.000000 & na & -0.0002 & 0.00 & -0.27 & -0.01 \\
0.000000 & na & -0.0005 & 0.00 & -0.01 & 0.03 \\
0.000000 & na & -0.001 & 0.00 & 0.01 & 0.00 \\
0.000000 & na & -0.002 & 0.00 & -0.01 & -0.01 \\
0.000000 & na & -0.005 & 0.00 & -0.01 & -0.01 \\
0.000000 & na & -0.01 & 0.00 & -0.02 & -0.02 \\
0.000625 & 0.001 & -0.0001 & 0.08 & -0.43 & -0.07 \\
0.000625 & 0.001 & -0.0002 & 0.10 & -0.21 & 0.06 \\
0.000625 & 0.001 & -0.0005 & 0.10 & 0.02 & 0.06 \\
0.000625 & 0.001 & -0.001 & 0.09 & 0.08 & 0.08 \\
0.000625 & 0.001 & -0.002 & 0.09 & 0.08 & 0.08 \\
0.000625 & 0.001 & -0.005 & 0.08 & 0.03 & 0.05 \\
0.000625 & 0.001 & -0.01 & 0.10 & 0.10 & 0.12 \\
0.000125 & 0.005 & -0.0001 & 0.09 & -0.46 & -0.06 \\
0.000125 & 0.005 & -0.0002 & 0.10 & -0.21 & 0.05 \\
0.000125 & 0.005 & -0.0005 & 0.10 & 0.05 & 0.13 \\
0.000125 & 0.005 & -0.001 & 0.10 & 0.06 & 0.07 \\
0.000125 & 0.005 & -0.002 & 0.10 & 0.10 & 0.10 \\
0.000125 & 0.005 & -0.005 & 0.08 & 0.14 & 0.13 \\
0.000125 & 0.005 & -0.01 & 0.10 & 0.12 & 0.14 \\
0.000063 & 0.01 & -0.0001 & 0.09 & -0.38 & -0.04 \\
0.000063 & 0.01 & -0.0002 & 0.09 & -0.20 & 0.09 \\
0.000063 & 0.01 & -0.0005 & 0.10 & 0.05 & 0.12 \\
0.000063 & 0.01 & -0.001 & 0.09 & 0.05 & 0.06 \\
0.000063 & 0.01 & -0.002 & 0.10 & 0.13 & 0.12 \\
0.000063 & 0.01 & -0.005 & 0.10 & 0.09 & 0.09 \\
0.000063 & 0.01 & -0.01 & 0.10 & 0.10 & 0.09 \\
0.000013 & 0.05 & -0.0001 & 0.07 & -0.35 & -0.03 \\
0.000013 & 0.05 & -0.0002 & 0.08 & -0.21 & 0.04 \\
0.000013 & 0.05 & -0.0005 & 0.09 & 0.04 & 0.11 \\
0.000013 & 0.05 & -0.001 & 0.09 & 0.10 & 0.13 \\
0.000013 & 0.05 & -0.002 & 0.09 & 0.12 & 0.12 \\
0.000013 & 0.05 & -0.005 & 0.10 & 0.07 & 0.09 \\
0.000013 & 0.05 & -0.01 & 0.09 & 0.09 & 0.10 \\
0.001250 & 0.001 & -0.0001 & 0.17 & -0.32 & 0.07 \\
0.001250 & 0.001 & -0.0002 & 0.19 & -0.14 & 0.14 \\
0.001250 & 0.001 & -0.0005 & 0.19 & 0.11 & 0.17 \\
0.001250 & 0.001 & -0.001 & 0.17 & 0.18 & 0.15 \\
0.001250 & 0.001 & -0.002 & 0.19 & 0.13 & 0.12 \\
0.001250 & 0.001 & -0.005 & 0.18 & 0.18 & 0.16 \\
0.001250 & 0.001 & -0.01 & 0.18 & 0.19 & 0.15 \\
0.000250 & 0.005 & -0.0001 & 0.18 & -0.34 & 0.04 \\
0.000250 & 0.005 & -0.0002 & 0.20 & -0.14 & 0.18 \\
0.000250 & 0.005 & -0.0005 & 0.20 & 0.12 & 0.15 \\
0.000250 & 0.005 & -0.001 & 0.19 & 0.18 & 0.17 \\
0.000250 & 0.005 & -0.002 & 0.19 & 0.20 & 0.20 \\
0.000250 & 0.005 & -0.005 & 0.18 & 0.17 & 0.18 \\
0.000250 & 0.005 & -0.01 & 0.18 & 0.18 & 0.20 \\
\end{tabular}
\newpage
\begin{tabular}{c|c|c|c|c|c}
$\rho_b$ & $s_b$ & $s_d$ & true $\alpha$ & MK $\alpha$ $(0.1)$ & MK $\alpha$ $(0.5)$\bigstrut \\
\hline
0.000125 & 0.01 & -0.0001 & 0.16 & -0.27 & 0.07 \\
0.000125 & 0.01 & -0.0002 & 0.18 & -0.13 & 0.16 \\
0.000125 & 0.01 & -0.0005 & 0.19 & 0.11 & 0.18 \\
0.000125 & 0.01 & -0.001 & 0.21 & 0.17 & 0.19 \\
0.000125 & 0.01 & -0.002 & 0.18 & 0.16 & 0.17 \\
0.000125 & 0.01 & -0.005 & 0.19 & 0.21 & 0.19 \\
0.000125 & 0.01 & -0.01 & 0.20 & 0.17 & 0.17 \\
0.000025 & 0.05 & -0.0001 & 0.13 & -0.16 & 0.04 \\
0.000025 & 0.05 & -0.0002 & 0.16 & -0.15 & 0.09 \\
0.000025 & 0.05 & -0.0005 & 0.18 & 0.06 & 0.19 \\
0.000025 & 0.05 & -0.001 & 0.18 & 0.18 & 0.20 \\
0.000025 & 0.05 & -0.002 & 0.18 & 0.19 & 0.19 \\
0.000025 & 0.05 & -0.005 & 0.19 & 0.20 & 0.17 \\
0.000025 & 0.05 & -0.01 & 0.18 & 0.18 & 0.16 \\
0.001875 & 0.001 & -0.0001 & 0.26 & -0.32 & 0.08 \\
0.001875 & 0.001 & -0.0002 & 0.28 & -0.09 & 0.23 \\
0.001875 & 0.001 & -0.0005 & 0.28 & 0.20 & 0.23 \\
0.001875 & 0.001 & -0.001 & 0.29 & 0.25 & 0.23 \\
0.001875 & 0.001 & -0.002 & 0.28 & 0.24 & 0.21 \\
0.001875 & 0.001 & -0.005 & 0.28 & 0.25 & 0.24 \\
0.001875 & 0.001 & -0.01 & 0.29 & 0.27 & 0.26 \\
0.000375 & 0.005 & -0.0001 & 0.24 & -0.26 & 0.12 \\
0.000375 & 0.005 & -0.0002 & 0.29 & -0.09 & 0.21 \\
0.000375 & 0.005 & -0.0005 & 0.29 & 0.22 & 0.26 \\
0.000375 & 0.005 & -0.001 & 0.29 & 0.30 & 0.27 \\
0.000375 & 0.005 & -0.002 & 0.29 & 0.27 & 0.28 \\
0.000375 & 0.005 & -0.005 & 0.29 & 0.26 & 0.24 \\
0.000375 & 0.005 & -0.01 & 0.26 & 0.30 & 0.25 \\
0.000188 & 0.01 & -0.0001 & 0.24 & -0.23 & 0.12 \\
0.000188 & 0.01 & -0.0002 & 0.27 & -0.06 & 0.22 \\
0.000188 & 0.01 & -0.0005 & 0.27 & 0.20 & 0.26 \\
0.000188 & 0.01 & -0.001 & 0.28 & 0.28 & 0.27 \\
0.000188 & 0.01 & -0.002 & 0.29 & 0.27 & 0.29 \\
0.000188 & 0.01 & -0.005 & 0.30 & 0.28 & 0.27 \\
0.000188 & 0.01 & -0.01 & 0.29 & 0.27 & 0.26 \\
0.000038 & 0.05 & -0.0001 & 0.17 & -0.05 & 0.14 \\
0.000038 & 0.05 & -0.0002 & 0.23 & -0.06 & 0.17 \\
0.000038 & 0.05 & -0.0005 & 0.28 & 0.15 & 0.26 \\
0.000038 & 0.05 & -0.001 & 0.28 & 0.27 & 0.28 \\
0.000038 & 0.05 & -0.002 & 0.28 & 0.27 & 0.27 \\
0.000038 & 0.05 & -0.005 & 0.28 & 0.30 & 0.31 \\
0.000038 & 0.05 & -0.01 & 0.30 & 0.31 & 0.31 \\
0.002500 & 0.001 & -0.0001 & 0.35 & -0.22 & 0.15 \\
0.002500 & 0.001 & -0.0002 & 0.40 & -0.02 & 0.29 \\
0.002500 & 0.001 & -0.0005 & 0.37 & 0.29 & 0.30 \\
0.002500 & 0.001 & -0.001 & 0.38 & 0.34 & 0.30 \\
0.002500 & 0.001 & -0.002 & 0.39 & 0.33 & 0.29 \\
0.002500 & 0.001 & -0.005 & 0.38 & 0.35 & 0.29 \\
0.002500 & 0.001 & -0.01 & 0.38 & 0.36 & 0.32 \\
\end{tabular}
\newpage
\begin{tabular}{c|c|c|c|c|c}
$\rho_b$ & $s_b$ & $s_d$ & true $\alpha$ & MK $\alpha$ $(0.1)$ & MK $\alpha$ $(0.5)$\bigstrut \\
\hline
0.000500 & 0.005 & -0.0001 & 0.30 & -0.18 & 0.21 \\
0.000500 & 0.005 & -0.0002 & 0.38 & 0.01 & 0.32 \\
0.000500 & 0.005 & -0.0005 & 0.39 & 0.29 & 0.37 \\
0.000500 & 0.005 & -0.001 & 0.39 & 0.38 & 0.36 \\
0.000500 & 0.005 & -0.002 & 0.39 & 0.36 & 0.35 \\
0.000500 & 0.005 & -0.005 & 0.38 & 0.37 & 0.39 \\
0.000500 & 0.005 & -0.01 & 0.37 & 0.37 & 0.37 \\
0.000250 & 0.01 & -0.0001 & 0.30 & -0.16 & 0.17 \\
0.000250 & 0.01 & -0.0002 & 0.34 & 0.04 & 0.34 \\
0.000250 & 0.01 & -0.0005 & 0.38 & 0.29 & 0.38 \\
0.000250 & 0.01 & -0.001 & 0.38 & 0.37 & 0.38 \\
0.000250 & 0.01 & -0.002 & 0.39 & 0.38 & 0.35 \\
0.000250 & 0.01 & -0.005 & 0.36 & 0.36 & 0.36 \\
0.000250 & 0.01 & -0.01 & 0.37 & 0.38 & 0.33 \\
0.000050 & 0.05 & -0.0001 & 0.21 & 0.01 & 0.15 \\
0.000050 & 0.05 & -0.0002 & 0.27 & 0.04 & 0.21 \\
0.000050 & 0.05 & -0.0005 & 0.35 & 0.20 & 0.28 \\
0.000050 & 0.05 & -0.001 & 0.38 & 0.33 & 0.39 \\
0.000050 & 0.05 & -0.002 & 0.38 & 0.39 & 0.39 \\
0.000050 & 0.05 & -0.005 & 0.37 & 0.38 & 0.37 \\
0.000050 & 0.05 & -0.01 & 0.36 & 0.40 & 0.36 \\
0.003125 & 0.001 & -0.0001 & 0.41 & -0.20 & 0.18 \\
0.003125 & 0.001 & -0.0002 & 0.48 & 0.04 & 0.36 \\
0.003125 & 0.001 & -0.0005 & 0.46 & 0.32 & 0.37 \\
0.003125 & 0.001 & -0.001 & 0.47 & 0.42 & 0.40 \\
0.003125 & 0.001 & -0.002 & 0.47 & 0.44 & 0.40 \\
0.003125 & 0.001 & -0.005 & 0.48 & 0.41 & 0.39 \\
0.003125 & 0.001 & -0.01 & 0.47 & 0.43 & 0.40 \\
0.000625 & 0.005 & -0.0001 & 0.39 & -0.11 & 0.25 \\
0.000625 & 0.005 & -0.0002 & 0.43 & 0.10 & 0.40 \\
0.000625 & 0.005 & -0.0005 & 0.48 & 0.38 & 0.47 \\
0.000625 & 0.005 & -0.001 & 0.49 & 0.46 & 0.46 \\
0.000625 & 0.005 & -0.002 & 0.50 & 0.47 & 0.47 \\
0.000625 & 0.005 & -0.005 & 0.48 & 0.43 & 0.44 \\
0.000625 & 0.005 & -0.01 & 0.47 & 0.48 & 0.45 \\
0.000313 & 0.01 & -0.0001 & 0.33 & -0.11 & 0.23 \\
0.000313 & 0.01 & -0.0002 & 0.42 & 0.08 & 0.36 \\
0.000313 & 0.01 & -0.0005 & 0.48 & 0.35 & 0.46 \\
0.000313 & 0.01 & -0.001 & 0.48 & 0.45 & 0.49 \\
0.000313 & 0.01 & -0.002 & 0.49 & 0.50 & 0.47 \\
0.000313 & 0.01 & -0.005 & 0.48 & 0.47 & 0.48 \\
0.000313 & 0.01 & -0.01 & 0.48 & 0.48 & 0.46 \\
0.000063 & 0.05 & -0.0001 & 0.25 & 0.06 & 0.18 \\
0.000063 & 0.05 & -0.0002 & 0.31 & 0.07 & 0.25 \\
0.000063 & 0.05 & -0.0005 & 0.42 & 0.28 & 0.41 \\
0.000063 & 0.05 & -0.001 & 0.47 & 0.40 & 0.46 \\
0.000063 & 0.05 & -0.002 & 0.49 & 0.47 & 0.47 \\
0.000063 & 0.05 & -0.005 & 0.47 & 0.46 & 0.50 \\
0.000063 & 0.05 & -0.01 & 0.47 & 0.48 & 0.46
\end{tabular}}

\vspace{1em}

\noindent DFE parameters, true values of $\alpha$, and MK estimates of $\alpha$ under the two cutoff frequencies 0.1 and 0.5 for all simulation runs from Figs. 1A and 2. The value of $\rho_b$ specifies the fraction of adaptive mutations among all functional mutations. 

\newpage

\subsection*{Figure S1}

\begin{figure}[h!]
\centering
\includegraphics[width=0.5\linewidth]{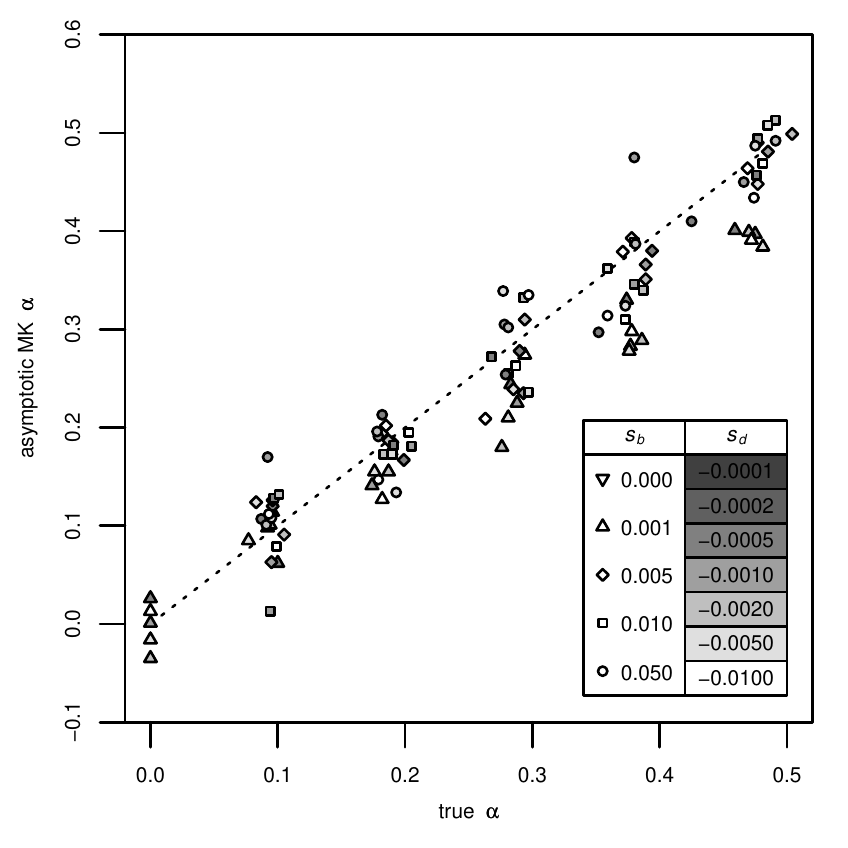}
\end{figure}

\noindent Comparison of true values of $\alpha$ and asymptotic MK estimates for all simulation runs from Table S1. The asymptotic MK estimates were obtained by fitting $\alpha(x)$ to an exponential function $\alpha(x)=a+b\exp{(-cx)}$ for all $x\ge 0.1$, using a nonlinear least-squares algorithm and extrapolating to $x=1$.

\end{document}